\documentclass[nocite]{epl}

\title{Precise measurements of UV atomic lines: Hyperfine \\
structure and isotope shifts in the 398.8 nm line of Yb}
\shorttitle{Precise measurements of UV atomic lines}
\author{Ayan Banerjee \and Umakant D. Rapol \and Dipankar 
Das \and Anusha Krishna \and \\ Vasant Natarajan
\thanks{E-mail: \email{vasant@physics.iisc.ernet.in}}}
\shortauthor{Ayan Banerjee \etal}
\institute{Department of Physics, Indian Institute of 
Science, Bangalore 560 012,  INDIA}
\pacs{32.30.Jc}{Visible and ultraviolet spectra}
\pacs{32.10.Fn}{Fine and hyperfine structure}
\pacs{42.62.Fi}{Laser spectroscopy}

\begin{document}
\maketitle

\begin{abstract}
We demonstrate a technique for frequency measurements of UV 
transitions with sub-MHz precision. The frequency is 
measured using a ring-cavity resonator whose length is 
calibrated against a reference laser locked to the $D_2$ 
line of $^{87}$Rb. We have used this to measure the 398.8 nm 
${^1S}_0 \leftrightarrow {^1P}_1$ line of atomic Yb. We 
report isotope shifts of all the seven stable isotopes, 
including the rarest isotope $^{168}$Yb. We have been able 
to resolve the overlapping $^{173}$Yb($F = 3/2$) and 
$^{172}$Yb transitions for the first time. We also obtain 
high-precision measurements of excited-state hyperfine 
structure in the odd isotopes, $^{171}$Yb and $^{173}$Yb. 
The measurements resolve several discrepancies among earlier 
measurements.
\end{abstract}

Precise measurements of the frequencies of atomic 
transitions are an important tool in expanding our knowledge 
of physics. For example, precise measurement of the $D_1$ 
line in Cs \cite{URH99} combined with an 
atom-interferometric measurement of the photon recoil shift 
\cite{WYC93} could lead to a more accurate determination of 
the fine-structure constant $\alpha$. In addition, 
hyperfine-structure and isotope-shift measurements in atomic 
lines can help in fine-tuning the atomic wavefunction, 
particularly due to contributions from nuclear interactions. 
This is important when comparing theoretical calculations 
with experimental data in atomic studies of parity violation 
\cite{WBC97}. The most precise optical frequency measurements to 
date have been done using the recently developed 
frequency-comb method with mode-locked lasers \cite{URH99}, with 
errors below 100 kHz being reported. However, to the best of 
our knowledge, this technique has not yet been applied to UV 
spectroscopy, which relies on older and less-accurate 
techniques.

In this Letter, we present the most comprehensive study of 
the 398.8 nm ${^1S}_0 \leftrightarrow {^1P}_1$ line of 
atomic Yb. Yb (Z=70) is an attractive candidate for studying 
atomic parity violation \cite{DEM95} and the search for a 
permanent electric-dipole moment in atoms \cite{HTK99}. 
Laser-cooled Yb has also been proposed for 
frequency-standards applications \cite{HZB89}. The ${^1S}_0 
\leftrightarrow {^1P}_1$ line is widely used in 
laser-cooling experiments \cite{HTK99,LBM01}. Over the years, 
there has been much interest in this line, and its isotopic 
and hyperfine components have been measured using a variety 
of techniques -- level-crossing and anti-crossing 
spectroscopy \cite{BUS69,BLL76,LIE85}, Fabry-Perot cavity 
\cite{GGA79}, saturated-absorption spectroscopy 
\cite{BEM92}, photon-burst spectroscopy \cite{DGK93}, and 
most recently using optical double-resonance spectroscopy 
with cold atoms in a magneto-optic trap \cite{LBM01}. 
However, all these measurements have errors of several MHz 
and show wide discrepancies with each other. We report the 
first sub-MHz measurement of the isotope shifts of all seven 
stable isotopes, including the rarest isotope $^{168}$Yb 
(natural abundance = 0.13\%). We have been able to resolve 
the overlapping $^{173}$Yb($F = 3/2$) and $^{172}$Yb 
transitions for the first time. We also report 
high-precision measurements of excited-state hyperfine structure 
in the odd isotopes, $^{171}$Yb and $^{173}$Yb.

Our novel technique for the frequency measurement uses the 
fact that the {\it absolute} frequency of the $D_2$ line 
($5S_{1/2} \leftrightarrow 5P_{3/2}$ transition) in 
$^{87}$Rb has been measured with an accuracy of 10 kHz 
\cite{YSJ96}. A stabilized diode laser locked to this line 
is used as a frequency reference along with a ring-cavity 
resonator whose length is locked to the reference laser. For 
a given cavity length, an unknown laser on an atomic 
transition has a small frequency offset from the nearest 
cavity resonance. This offset is combined with the cavity 
mode number to obtain a precise value for the {\it absolute} 
frequency of 
the unknown laser. We have earlier used this technique to 
make measurements of hyperfine intervals in the 
$D_2$ line of $^{85}$Rb with 30 kHz precision \cite{BDN03}, 
where we have also 
highlighted the advantages of the Rb-stabilized ring cavity. 
In this work, we apply the technique for UV measurements 
using a frequency-doubled IR laser to access the UV lines. 
By measuring the frequency of the IR laser and not the UV 
laser, we avoid several complications associated with UV 
spectroscopy. The technique is uniquely suited for measuring 
hyperfine intervals and isotope shifts since several sources 
of systematic error cancel in such measurements.

\begin{figure}
\twoimages[scale=0.45]{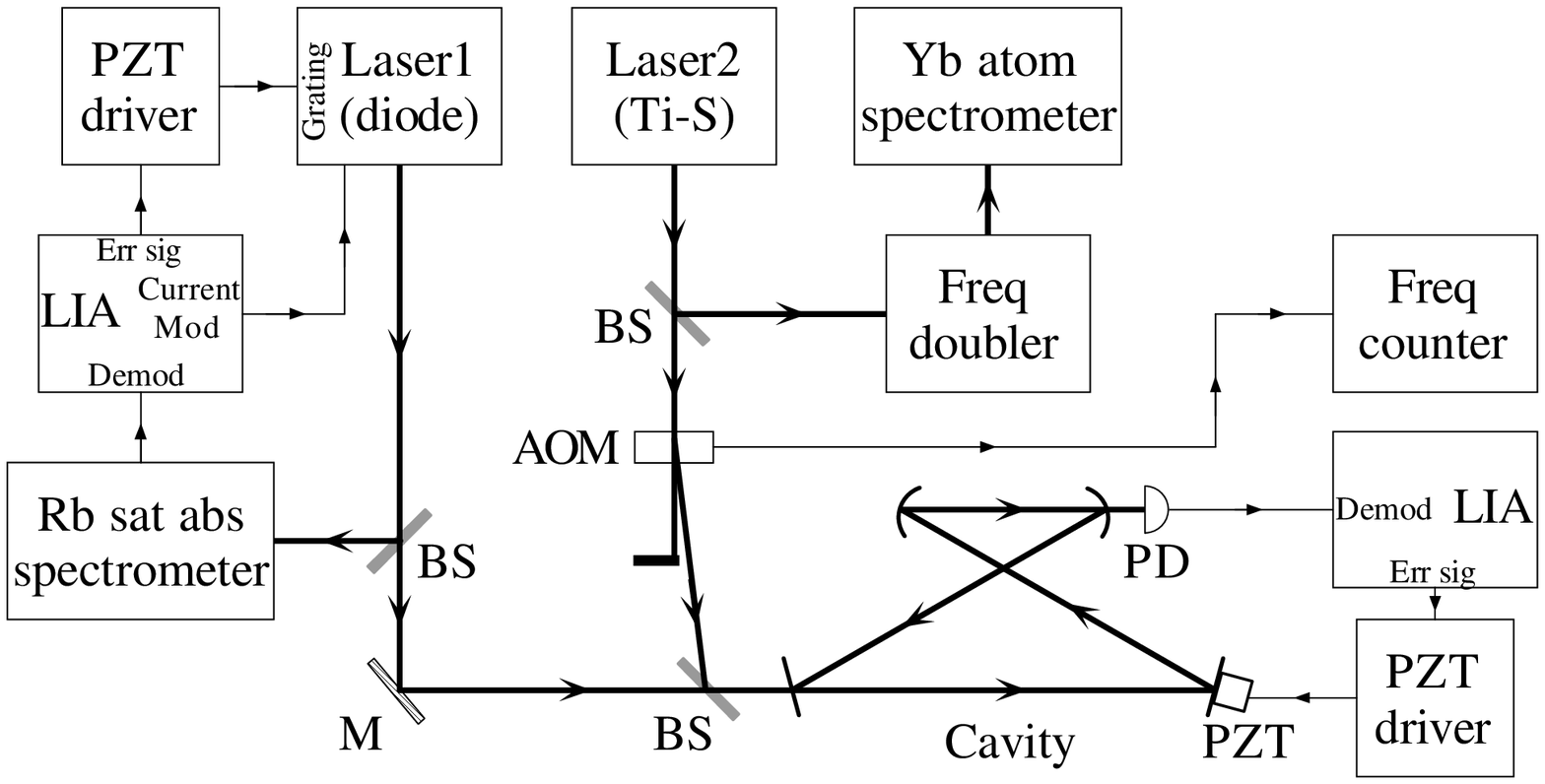}{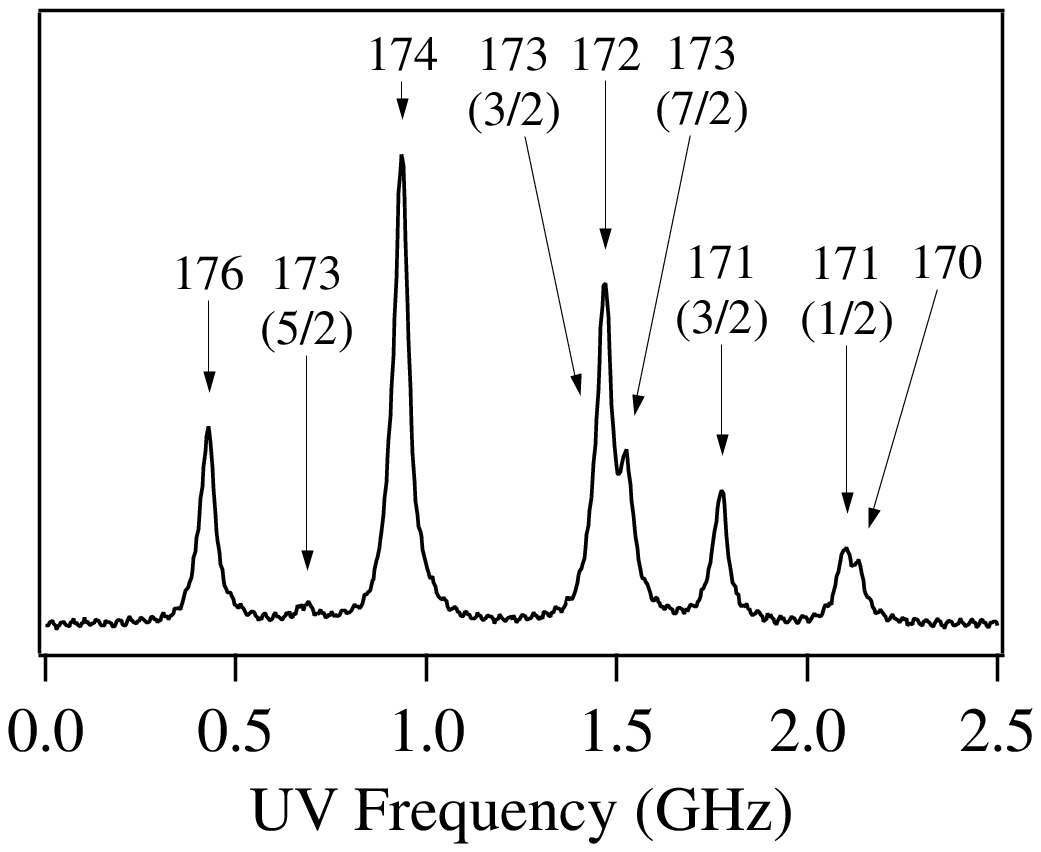}
\caption{
Schematic of the experiment. The trace on the right is the 
fluorescence signal showing the different Yb transitions. 
Two peaks contain overlapping transitions and need to be 
separated by multi-peak fitting.}
\label{f1}
\end{figure}

The schematic of the experiment is shown in Fig.\ \ref{f1}. 
Laser1 is a frequency-stabilized diode laser that acts as 
the frequency reference for the cavity. It is locked to the 
$D_2$ line of $^{87}$Rb using saturated-absorption 
spectroscopy in a vapor cell. Laser2 is a tunable 
Ti-sapphire laser (Coherent 899-21) whose output is fed 
into an external frequency doubler (Laser Analytical Systems
LAS100) to access the 398.8 nm Yb line. The doubler uses a 
patented $\Delta$-cavity design that allows it to track large
frequency scans of the Ti-S laser (up to 30 GHz) without 
mode hops. The Ti-S laser is frequency stabilized to a 
linewidth of 500 kHz. A part of its output is tapped off 
before the doubler and coupled into the 
ring-cavity resonator for the frequency measurement. Yb 
spectroscopy is done inside a vacuum chamber maintained at a 
pressure below $10^{-9}$ torr. The Yb atomic beam is 
produced by heating a quartz ampoule containing metallic Yb 
to a temperature of 400$^\circ$C. To reduce Doppler 
broadening, the laser beam intersects the atomic beam at 
right angles and the fluorescence from the atoms is detected 
through a narrow slit. A typical Yb spectrum is shown in 
Fig.\ \ref{f1}b. With 20 mW of UV light in a beam of 
diameter 1 cm, we achieve a linewidth of 45 MHz, or 
about 50\% larger than the natural linewidth of 28 MHz.

\begin{figure}
\twofigures[scale=0.4]{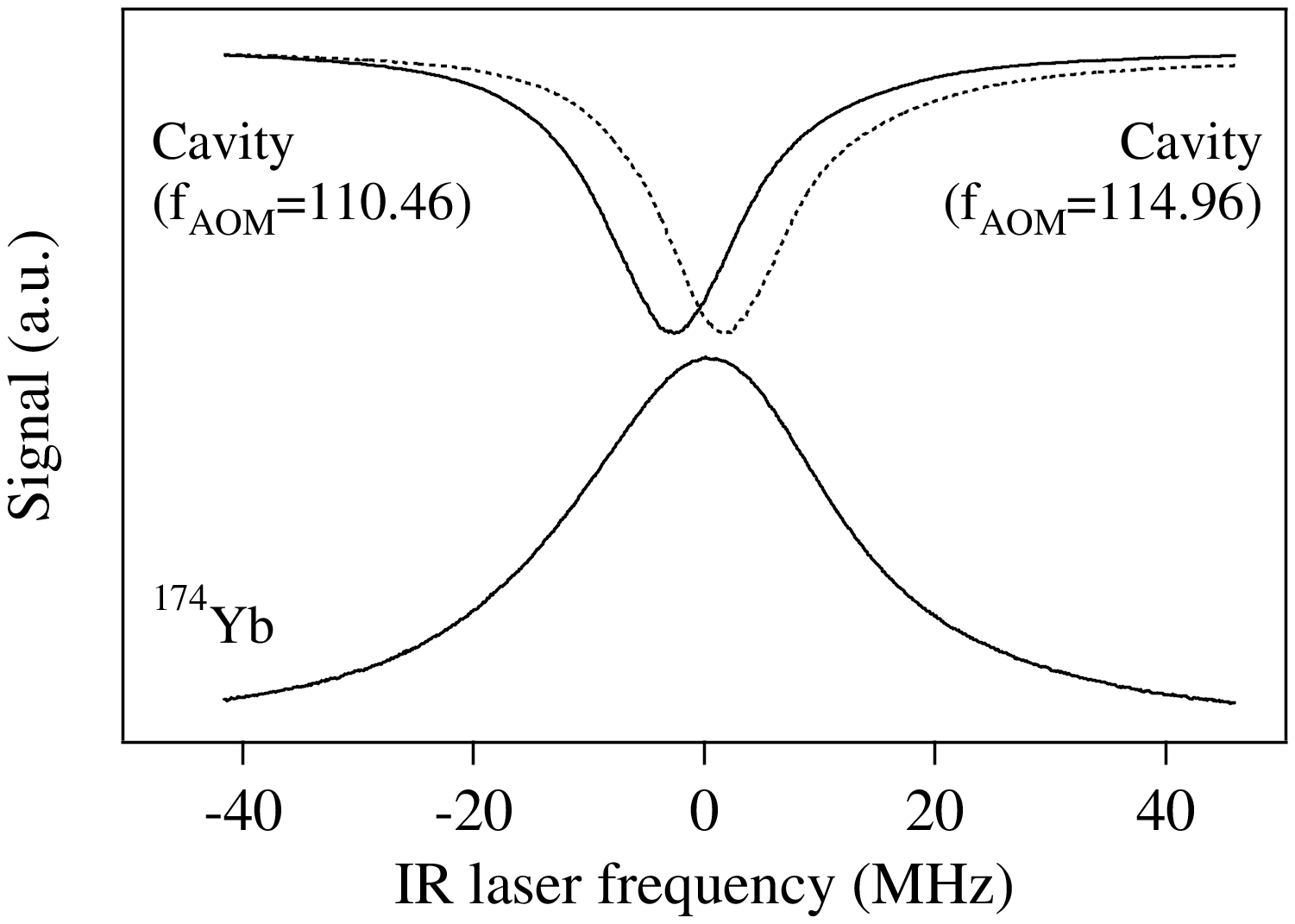}{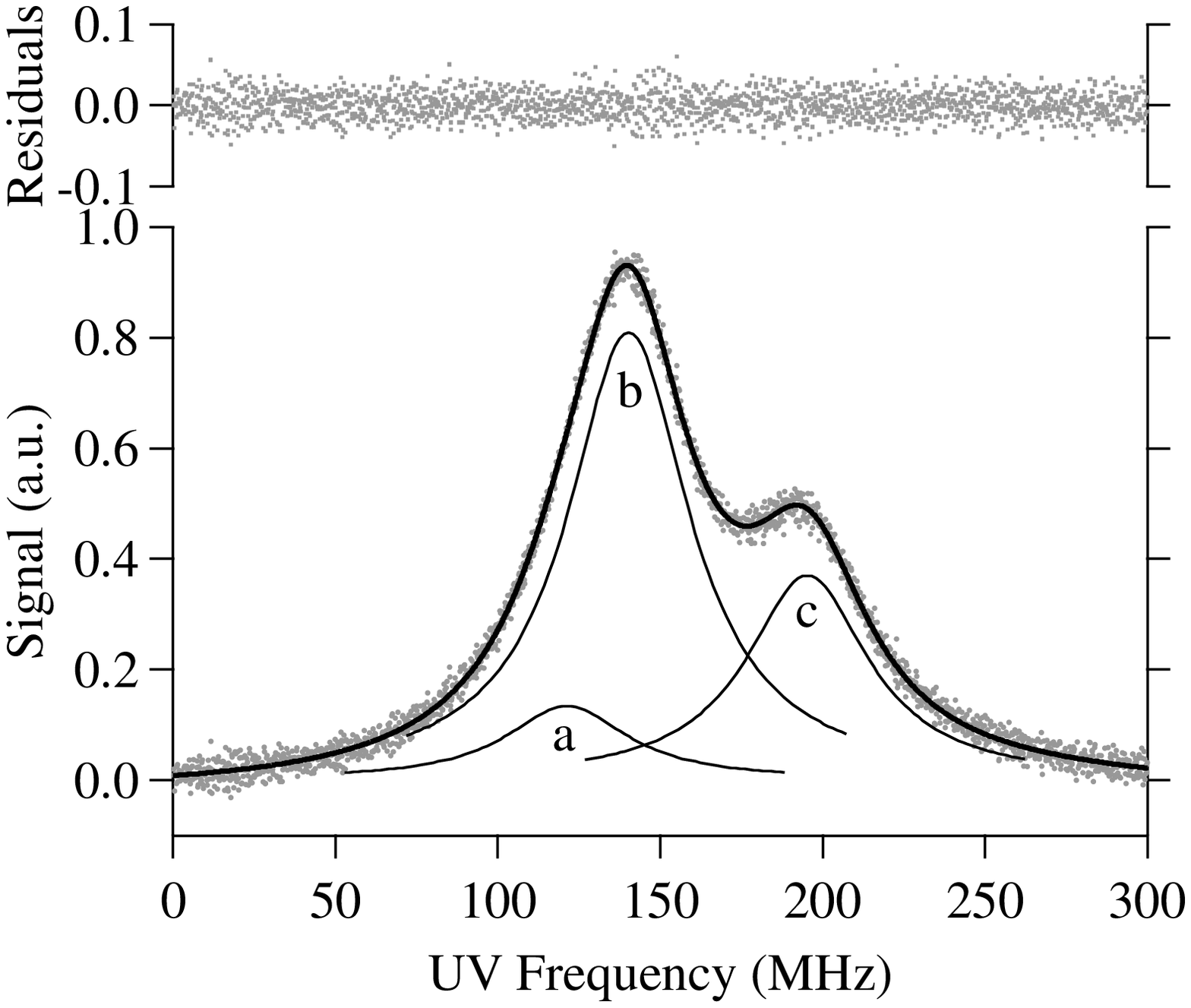}
\caption{Determination of the frequency offset. The lower trace is 
the fluorescence signal as the laser is scanned across the 
$^{174}$Yb transition. The upper traces are the 
cavity-reflection signal for two values of the AOM frequency: 
110.46 MHz for the solid line and 114.96 MHz for the dotted 
line. A fit to the separation between the fluorescence and 
cavity peaks vs.\ the AOM frequency yields the exact 
frequency that brings the cavity on resonance, which in this 
case is 112.64 MHz.}
\label{f2}
\caption{
Three-peak fitting to extract overlapping transitions. The 
fluorescence signal (shown as gray circles) is fitted to 
three Lorentzians with the same linewidths. The three peaks 
correspond to $^{173}$Yb$(F=3/2)$ labeled as a, $^{172}$Yb 
labeled as b, and $^{173}$Yb$(F=7/2)$ labeled as c. The 
thick line is the sum of the three peaks and fits the 
measured data very well as seen from the structure-less 
residuals shown on top.}
\label{f3}
\end{figure}

The frequency measurement proceeds as follows. The outputs 
of Laser1 (reference) and Laser2 (Ti-S) are fed into the 
ring cavity. The cavity length is adjusted using a 
piezo-mounted mirror to bring it into resonance with the 
wavelength of Laser1. The cavity is then locked to this 
length in a feedback loop. However, Laser2 will still be 
offset from the cavity resonance. This offset is accounted 
for by shifting the frequency of the laser using an 
acousto-optic modulator (AOM) before it enters the cavity. The 
signal from a scan of Laser2 is shown in Fig.\ \ref{f2}. The 
upper trace shows the reflected signal from the cavity, 
which goes to a minimum as the cavity comes into resonance. 
The lower trace is the fluorescence signal corresponding to 
the $^{174}$Yb transition. The cavity resonance slightly to 
the left of the $^{174}$Yb peak (solid line) is for an AOM 
frequency of 110.46 MHz, while the one to the right (dotted 
line) is for a frequency of 114.96 MHz. Thus, by changing 
the AOM frequency, we can move the cavity resonance across 
the $^{174}$Yb peak. The peak centers are determined to an accuracy
of 50 kHz by fitting a Lorentzian lineshape, and we
verify that there is no significant deviation from this lineshape 
from the featureless residuals. 
A straight-line fit to the peak separation 
vs.\ the AOM frequency gives the AOM 
frequency that brings the cavity into resonance (with a typical
accuracy of 100 kHz). Once the exact 
cavity length (or mode number) is known, the absolute frequency 
of Laser1 is used to determine the absolute frequency of Laser2.

The measurements rely on the fact that the cavity mode 
number is known exactly. For this, we measure the cavity 
free-spectral range (fsr) very precisely, in the following 
manner. We first lock the cavity with the reference laser on 
the $F = 2 \rightarrow F' = (2,3)$ 
transition in $^{87}$Rb and measure 
the AOM offset for the $^{174}$Yb transition. We then shift 
the reference laser to the $F = 1 \rightarrow F' = (1,2)$ 
transition, which is exactly 6622.887 MHz higher 
\cite{AIV77,YSJ96}. This shift causes the cavity mode number 
to increase by almost exactly 5 since the fsr is about 1326 
MHz. The cavity is locked to the new frequency and the AOM 
offset for the same $^{174}$Yb transition is measured. The 
difference between the two AOM offsets along with the change in 
the reference frequency gives exactly 5 times the cavity 
fsr. Using this method, we determine the fsr with a 
precision of 30 kHz. To determine the mode number, we 
measure the frequency of the $^{174}$Yb transition to an 
accuracy of 20 MHz using a home-built wavemeter 
\cite{BRW01}. Thus, there is a {\it unique} mode-number that 
matches the cavity resonance condition and the measured fsr. 
Indeed, the next nearest mode that satisfies the resonance 
condition has an fsr differing by 625 kHz, or about 20 times 
the error in the determination of the fsr. Similarly, a 
change in the mode number by one causes the frequency of the 
$^{174}$Yb transition to change by 60 MHz, or about 3 times 
the error with which the frequency is known.

The Yb spectrum has two peaks where there is significant 
overlap between neighboring transitions. One of these 
consists of the $^{171}$Yb($F=1/2$) and the $^{170}$Yb 
transitions. This does not present any problem in our 
technique because the two transitions are separated by about 
40 MHz and the spectrum shows two clear maxima. A 
two-Lorentzian fitting algorithm uniquely extracts the locations 
of the two transitions. However, the other multiple peak 
consists of the $^{173}$Yb($F=3/2$), $^{172}$Yb, and 
$^{173}$Yb($F = 7/2$) transitions, all lying within 50 MHz 
of each other. As seen from Fig.\ \ref{f3}, the first two 
transitions have a separation less than the natural 
linewidth and are completely merged. When we fit a single 
Lorentzian to these two overlapping transitions, we find its 
linewidth to be 1.4 times larger than the linewidth of the 
neighboring $^{173}$Yb($F = 7/2$) transition, and indeed the 
linewidth of all the other peaks in the spectrum. This gives 
us confidence that the increased linewidth is a result of 
the convolution of two individual Lorentzians. Therefore, we 
fit two Lorentzians to the overlapping peaks, and a third 
Lorentzian to the neighboring peak, with the 
constraint that the linewidth of all three peaks is the same. The 
algorithm then returns three peaks having linewidths similar 
to what we obtain for other well-resolved peaks. As seen 
from Fig.\ \ref{f3}, the fit residuals are very small and 
their structure-less noise shows that there is no ambiguity 
in the fitting. After fitting to about 30 spectra we obtain 
an average value of $17.64 \pm 0.90$ MHz for the separation 
between the $^{173}$Yb($F = 3/2$) and the $^{172}$Yb line 
centers. We believe this is the first direct measurement of 
this separation.

The first source of systematic error we consider arises due 
to improper perpendicular alignment of the laser beam with 
the Yb atomic beam. A misalignment angle of 10 mrad can 
cause a Doppler shift of about 7 MHz. To minimize this, we 
have repeated each measurement with a counter-propagating 
laser beam. Since the shift in this case is of opposite 
sign, the error cancels when we take an average of the two 
values. Indeed, the difference between the two values gives 
an estimate of the misalignment angle, which in our case is 
less than 2 mrad. In any case, the error cancels in the 
determination of the isotope shifts (relative to $^{174}$Yb) 
since all isotopes experience the same shift. Of course, 
there is a small differential Doppler shift due to the fact 
that the different isotopes leave the oven with slightly 
different velocities. But even for a large misalignment 
angle of 10 mrad the differential shift is only 120 kHz, 
which is negligible at our level of precision.

There are two classes of systematic error inherent to our 
technique. The first depends on variations of the reference 
laser. The possible causes are shifts in the laser lock 
point due to residual Doppler profile or optical-pumping 
effects in the Rb saturated-absorption spectrometer which 
change the lineshape of the peaks. We have tried to minimize 
this by using third-harmonic locking and careful control of 
pump and probe intensities in the spectrometer. Collisional 
shifts in the Rb vapor cell are estimated to be less than 10 
kHz. The magnetic field in the vicinity of the cell is less 
than 0.1 mT. Such magnetic fields cause broadening of the 
lines but do not shift the line center. As a further check 
on these errors, we have repeated the experiments with the 
vapor cell at different locations, and using vapor cells 
from different manufacturers. In our earlier work \cite{BDN03},
we have shown that systematic shifts in the reference frequency
are below 30 kHz. Finally, we have measured the 
same Yb transition using different hyperfine transitions to 
lock the reference laser. The reference frequency changes 
by known amounts (up to several GHz), and we have verified 
that the measured frequencies are consistent within the 
error bars. Changing the lock point of the reference laser
also changes the cavity mode and the AOM offset needed to bring 
the Yb transition into resonance with the cavity. This checks 
for systematic errors that might arise from variations in the 
direction of the beam entering the cavity when the AOM frequency
is varied.

The second class of systematic errors in our technique is 
wavelength dependent. Since our cavity is comparing the {\it 
wavelengths} of the two lasers, we have to convert these to 
frequencies using the refractive index of air \cite{EDL66}. 
Any error in the refractive index would reflect as a 
systematic shift in the measured frequency. From the 
reliability of the refractive index formulae, we estimate 
that this could cause a shift as large as 3 MHz in the 
measured frequencies. The only way to reduce this is to use 
an evacuated cavity, which we plan to do in the future. 
Wavelength-dependent systematic errors can also arise from 
varying phase shifts in the cavity mirrors, but we expect 
these to be negligible at the MHz level. It is important to 
note that both classes of systematic errors do not affect 
the determination of {\it frequency differences} of the 
unknown laser, up to several 10s of GHz. For example, in the 
measurement of the Yb line, these errors would cause all the 
frequencies to shift, but the {\it differences} listed in 
Table \ref{t1} will not change down to the kHz level. 

\begin{table}
\caption{ 
Listed are the various transitions of the 398.8 nm ${^1S}_0 
\leftrightarrow {^1P}_1$ line in Yb. The shifts from the 
$^{174}$Yb transition measured in this work are compared to 
values reported in earlier work. }
\begin{largetabular}{lrrrrr}
\hline\hline
& \multicolumn{5}{c}{Shift from $^{174}$Yb (MHz)} \\
\cline{2-6}
\multicolumn{1}{c}{Isotope} &\multicolumn{1}{c}{This work} & 
\multicolumn{1}{c}{Ref.\ \cite{LBM01}} 
&\multicolumn{1}{c}{Ref.\ \cite{DGK93}} 
&\multicolumn{1}{c}{Ref.\ \cite{GGA79}} 
&\multicolumn{1}{c}{Ref.\ \cite{CHA66}} \\
\hline
$^{176}$Yb & -509.98 $\pm$ 0.75 & -507.2 $\pm$ 2.5 &  & 
-509.4 $\pm$ 4.0 & -469.2 $\pm$ 2.7 \\
$^{173}$Yb ($F=5/2$) & -254.67 $\pm$ 0.63 &  &  &  & \\
$^{173}$Yb ($F=3/2$) & 516.26 $\pm$ 0.90 &  &  &  & \\
$^{172}$Yb  & 533.90 $\pm$ 0.70 & 527.8 $\pm$ 2.8 &  & 529.9 
$\pm$ 4.0 & 530.20 $\pm$ 7.80\\
$^{173}$Yb ($F=7/2$) & 589.00 $\pm$ 0.45 & 578.1 $\pm$ 5.8 &  
&  & \\
$^{171}$Yb ($F=3/2$) & 833.24 $\pm$ 0.75 & 832.5 $\pm$ 5.6 & 
834.4 $\pm$ 4.0 &  & \\
$^{171}$Yb ($F=1/2$) & 1152.86 $\pm$ 0.60 & 1151.4 $\pm$ 5.6 
& 1136.2 $\pm$ 5.8 &  & \\
$^{170}$Yb & 1192.48 $\pm$ 0.90 & 1175.7 $\pm$ 8.1 & 1172.5 
$\pm$ 5.7 & 1195.0 $\pm$ 10.8 & 1158.9 $\pm$ 11.4 \\
$^{168}$Yb & 1886.57 $\pm$ 1.00 &  & 1870.2 $\pm$ 5.2 &  &  
\\
$^{173}$Yb (centroid) & 291.61 $\pm$ 0.35 &  &  & 291.2 
$\pm$ 10.0 & \\
$^{171}$Yb (centroid) & 939.78 $\pm$ 0.54 & 938.8 $\pm$ 4.2 
& 935.0 $\pm$ 3.3 & 943.7 $\pm$ 7.0 & \\
\hline\hline
\end{largetabular}
\label{t1}
\end{table}

The measured frequencies of the various transitions in the 
${^1S}_0 \leftrightarrow {^1P}_1$ line of Yb are listed in 
Table \ref{t1}. Each value is an average of 6 individual 
measurements and the error quoted is the statistical error 
in the average. Even though we measure the absolute frequency 
of each transition, we have listed the values as shifts from 
the frequency of the $^{174}$Yb transition. This is because, 
as mentioned earlier, the measured frequencies 
may have large systematic errors due to the uncertainty 
in the refractive index, but the frequency differences have 
negligible error. Furthermore, this way of presenting the data 
allows direct comparison with previous work where only 
the shifts were measured. It is clear from the several blank 
entries for the previous work that our work is the most 
comprehensive measurement of this line to date. In addition, 
some of the previous results have non-overlapping error bars 
to the extent of $10 \sigma$ and our work resolves these 
discrepancies. For $^{176}$Yb, it is now accepted that the 
low value obtained by Chaiko \cite{CHA66} is 
incorrect. For $^{171}$Yb$(F=1/2)$, our value is in 
agreement with the recent value of 1151.4 MHz reported by 
Loftus et al.\ \cite{LBM01} but inconsistent with the value of 
1136.2 MHz reported by Deilamian et al.\ \cite{DGK93}. 
Indeed, the main result in the work of Deilamian et al.\ is to 
resolve the overlapping $^{171}$Yb$(F=1/2)$ and $^{170}$Yb 
transitions. The separation they measure is $36.3 \pm 2.5$ 
MHz, which is in good agreement with our value of $39.5 \pm 
0.4$ MHz. For $^{170}$Yb, the most discrepant value is the
result of Chaiko, where we have 
already seen that the value for $^{176}$Yb is incorrect. The 
most recent result for $^{170}$Yb is consistent only at the
$2 \sigma$ level, suggesting the need
for further high-precision measurements to verify our results. 
All the other shifts reported by us agree with previous work.

\begin{figure}
\twofigures[scale=0.45]{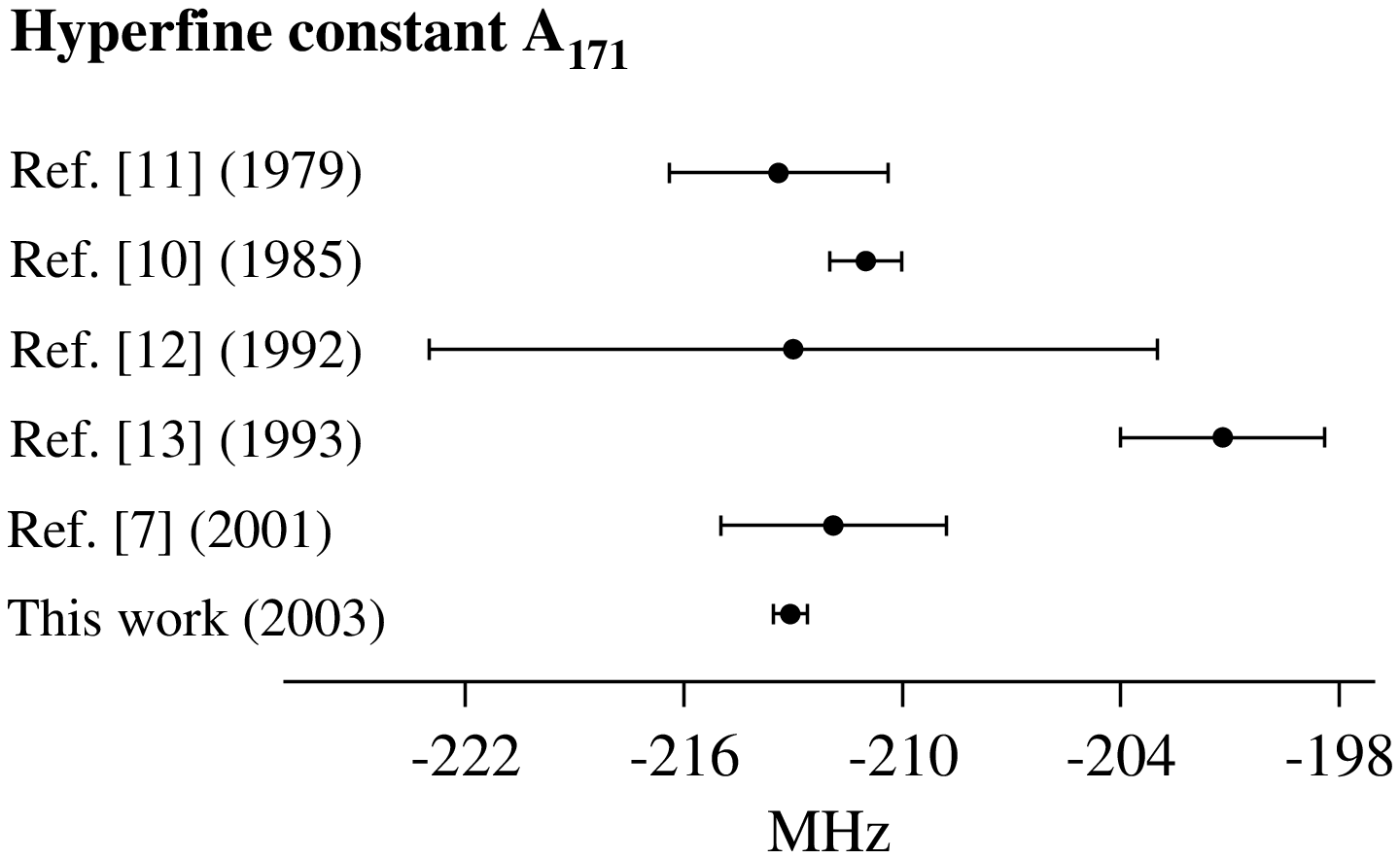}{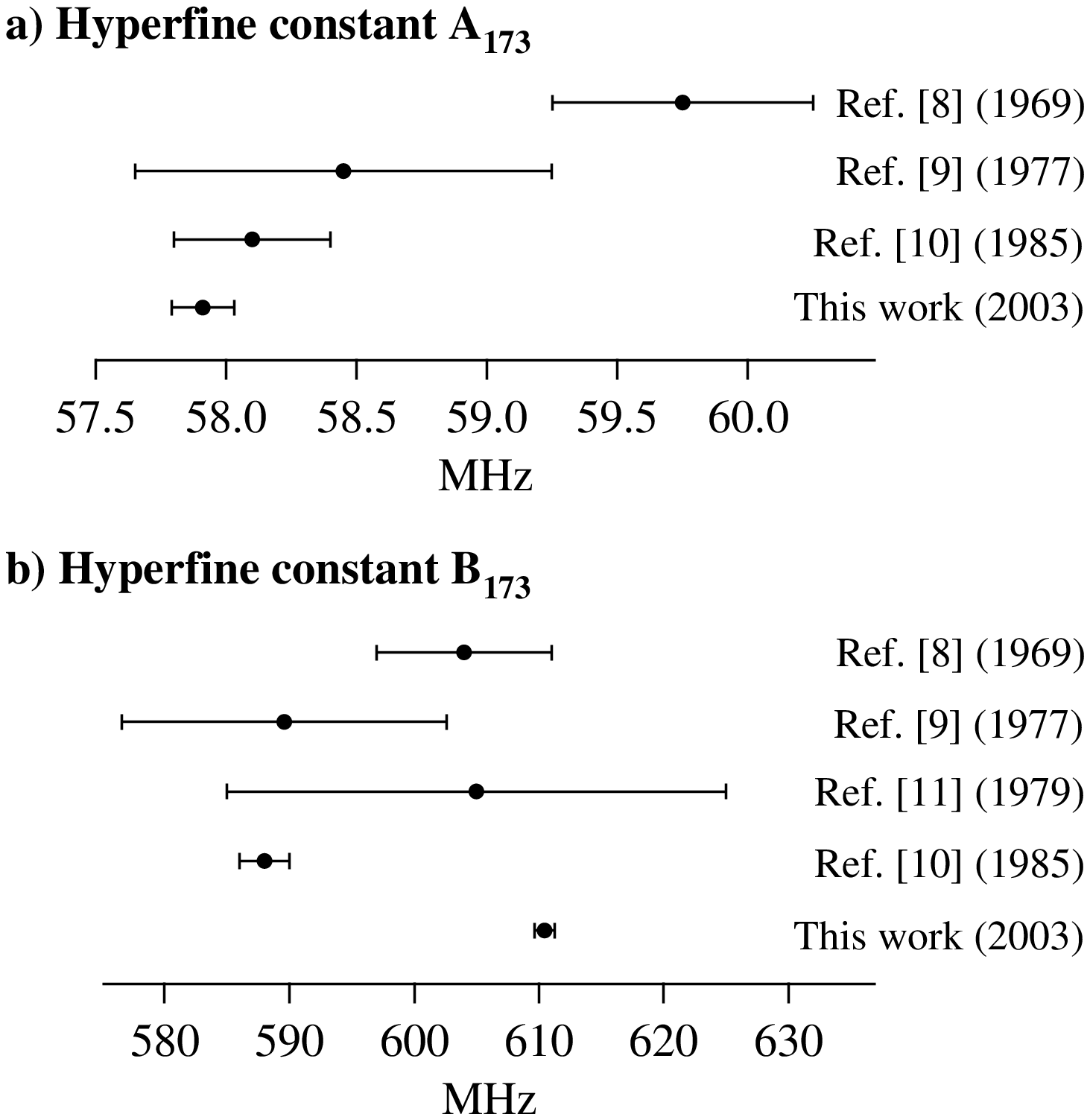}
\caption{
Hyperfine structure in $^{171}$Yb. The figure shows a 
comparison of our value of the hyperfine constant $A_{171}$ 
with earlier values.}
\label{f4}
\caption{
Hyperfine structure in $^{173}$Yb. The figure shows a 
comparison of our values of the hyperfine constants with 
earlier values: of $A_{173}$ in a) and of $B_{173}$ in b).}
\label{f5}
\end{figure}

We have used the data in Table \ref{t1} to obtain the 
hyperfine-coupling constants in the $6s6p$~${^1P}_1$ state 
of the odd isotopes, $^{171}$Yb and $^{173}$Yb. For 
$^{171}$Yb, the measured $\{ 3/2-1/2 \}$ interval is used to 
calculate the magnetic-dipole coupling constant $A$, 
yielding a value of $A_{171} = -213.08(47)$ MHz. This value 
is compared with earlier values in Fig.\ \ref{f4}. Our value 
is consistent but has significantly smaller error. For 
$^{173}$Yb, there are three hyperfine levels and the 
measured intervals are used to calculate the magnetic-dipole 
coupling constant $A$ and the electric-quadrupole coupling 
constant $B$. We obtain values of $A_{173}=57.91(12)$ MHz 
and $B_{173}=610.47(84)$ MHz. These values are again 
compared to earlier results in Figs.\ \ref{f5}a and b, 
respectively. While our value of $A_{173}$ is consistent, 
the value of $B_{173}$ is higher than the most recent value 
reported by Liening \cite{LIE85} from a level-crossing 
experiment. However, our value is in agreement with the 
other values. In addition, we have calculated the value of 
$g_J$ from the value of $A_{171}/g_J$ = 206.0(16) reported 
by Budick and Snir \cite{BUS69} using a level-crossing 
experiment. We obtain a value of $g_J=1.034(7)$, which is 
consistent with the currently accepted value of $1.035(5)$ 
\cite{GGA79}, giving further confidence in our 
results.

In conclusion, we have demonstrated a novel technique for 
measuring the frequencies of UV transitions 
using a Rb-stabilized ring-cavity resonator. The UV lines 
are accessed using a frequency-doubled IR laser. By 
measuring the frequency of the IR laser instead of the UV 
laser, we simplify the measurement process and avoid several 
sources of systematic errors. We have used this technique to 
measure various transitions in the 398.8 nm line in Yb. We 
obtain sub-MHz accuracy in the determination of isotope 
shifts for all isotopes (including the rarest isotope 
$^{168}$Yb), representing an order of magnitude or more 
improvement in precision. We have also determined hyperfine 
structure for the odd isotopes with high precision. Our 
technique actually gives the {\it absolute} frequencies of the 
different transitions. However, we have quoted only the 
shifts from the $^{174}$Yb transition since uncertainties in 
the refractive index of air in the cavity can cause an error 
as large as 3 MHz. In future, we plan to evacuate the cavity 
which will eliminate the need for refractive index 
correction. In addition, we plan to use a frequency-doubled 
{\it diode laser} for the UV spectroscopy. The output of the 
diode laser can be easily frequency modulated by modulating 
the injection current. The error signal generated can be fed 
back to lock the AOM frequency to a given peak, which will 
give us much higher accuracy in the frequency measurement. 
Using such a current-modulated diode laser, we have already 
demonstrated a precision of 30 kHz in measuring the 780 nm 
$D_2$ line of $^{85}$Rb \cite{BDN03}, and we hope to achieve 
similar precision in the measurement of UV lines.

This work was supported by the Board of Research in Nuclear 
Sciences (DAE), and the Department of Science and 
Technology, Government of India.

\end{document}